\documentclass{article}
\usepackage{amsmath}
\usepackage{amssymb}
\usepackage{amsfonts}
\usepackage{amsthm}
\usepackage{graphicx}
\usepackage{listings}

\newcommand\pdt[2]{\frac{\partial{#1}}{\partial{#2}}}		% Traditional way of writing partial derivatives
\newcommand\td[2]{\frac{d{#1}}{d{#2}}}					% Total derivative of order 1
\newcommand\etal{\mbox{\textit{et al.}}}
\newcommand{\uvec}[1]{\hat{e}_{#1}}

\newcommand{\grad}[1]{{\nabla}{#1}}
\newcommand{\dive}[1]{{\nabla}\cdot{#1}}
\newcommand{\curl}[1]{{\nabla}\times{#1}}
\newcommand{\Laplacian}[1]{\nabla^2{#1}}

\begin{document}
\title{Convection triggered by an electric field in a fluid heated from above}
\author{Amey Joshi\\Tata Consultancy Services,\\International Technology Park,\\Whitefield, Bangalore, 560 066\\}
\maketitle

%Abstract is required.
\begin{abstract}
I consider a dielectric fluid heated from above and subjected to an electric potential difference between its top and bottom. I show that for a suitably chosen electric potential 
difference, the layer of fluid can become unstable. For the case of a strongly polar fluid like pure water, an electric potential difference of a few hundreds of volts can trigger 
convection. Although the analysis in this paper cannot explain the phenomenon described in \cite{gp66}, it could be because of unavailability of accurate physical parameters of the 
fluid used in the experiment.
\end{abstract}

\section{Introduction}\label{intro}
A layer of insulating fluid is observed to develop a tessellated pattern of motions \cite{gp66} when it is heated from above and a sufficiently strong electric field is applied in a
direction parallel to gravity. The tessellations are similar to those observed in Rayleigh-B\`{e}nard convection. P. H. Roberts \cite{phr69} used linear analysis with Boussinesq
approximation, to estimate the voltage necessary to trigger convection. It was several orders of magnitude higher than the one observed in experiment. The same paper checked if free
charges induced in the fluid layer caused convection and found a negative answer. Since then, several authors have tried to explain why electric field triggers convection in an
arrangement that is otherwise stable. Physical mechanisms like temperature dependence of conductivity and permittivity \cite{ta} \cite{mr} and conductivity due to impurities \cite{wr}
\cite{rl} can neither satisfactorily explain the phenomenon nor correctly predict the voltage that triggers it \cite{bs}.

In this paper, I show that a temperature gradient in the fluid causes a variation in the electric field and the electric permittivity. Non-uniformities in the latter two quantities
result in a volume force, called Korteweg-Helmholtz force, to act on the fluid. Further, I use an alternative expression for the Korteweg-Helmholtz force that takes into account the 
dependence of electric field on mass density. It depends on the voltage applied across the plates and the vertical position in the fluid. A sufficiently
high voltage results in a distribution of the net force in such a manner as to trigger convection. Electrohydrodynamic (EHD) convection can thus be explained solely in terms of the volume
forces acting on the fluid. Further, the force depends only on the dielectric nature of the fluid and not its deviation from it.

In order to calculate the force density in a heated fluid, I begin with solving the electrostatic problem, of finding the electric potential in an infinite plane capacitor containing a
fluid dielectric, in the next section. I use its solution to find the electric field in the fluid. In section \ref{sec:vfd}, I derive force density in the fluid using the expressions for 
the electric field and the electric permittivity. The force density is finally used to derive a criterion for the onset of EHD convection in section \ref{sec:crit}. In section 
\ref{sec:veri}, I use the results to estimate a range of voltages that can trigger convection in water and transformer oil. The appendix has R code to generate plots shown in section 
\ref{sec:veri}.

\section{Solution of the electrostatic problem}\label{sec:electro}
\begin{figure}
\centering
\centerline{\includegraphics[scale=0.5]{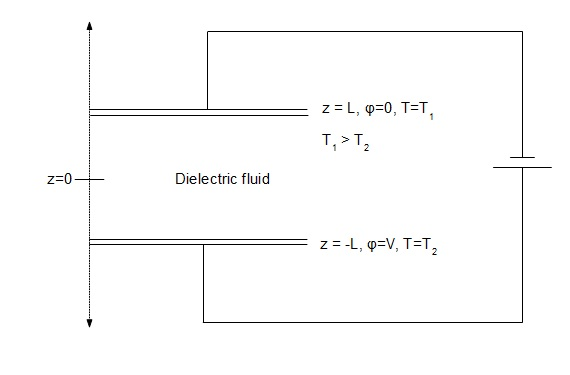}}
\caption{Experimental set up of Gross \etal}
\label{electro:f1}
\end{figure}

The schematic of the electro-hydrodynamic convection experiment used by Gross \etal \cite{gp66} is shown in figure \ref{electro:f1}. It consists of a capacitor cell having two rigid, 
conducting plates and containing a dielectric fluid. The top plate is fitted with a heating device. The plates are connected to a voltage source. If the heater is not switched on, the 
temperature is constant throughout the fluid and the voltage across the plates generates a uniform electric field. Since the electric permittivity is a function of temperature, turning on 
the heater sets up a permittivity gradient in addition to a temperature gradient in the fluid. The two gradients are parallel to each other. A non-uniform permittivity also makes the 
electric field to vary along the vertical direction. Gradients of permittivity and electric field result in a non-uniform volume force in the fluid making it unstable to perturbations.

For sake of simplicity, I assume that the plates are infinitely large and are defined by the equations $z=-L$ and $z=L$ respectively. I further assume that the permittivity is a linear 
function of temperature. Since the fluid is a dielectric, I first find the electric displacement $\vec{D}(\vec{x})$. It is given by Gauss' law $\dive{\vec{D}}=\rho_f(\vec{x})$, where 
$\rho_f(\vec{x})$ is the density of free charges and $\vec{x}$ denotes the position vector. Since there are no free charges in an ideal dielectric, 
\begin{equation}\label{e1}
\dive{\vec{D}} = 0
\end{equation}
Electric displacement is related to electric field $\vec{E}$ as $\vec{D}=\epsilon(T)\vec{E}$, where $\epsilon$ is the electric permittivity and $T$ is the absolute temperature. $T$ is a 
function of $z$ alone. Therefore $\epsilon$ can be considered to be a function of $z$ and
\begin{equation}\label{e2}
\curl{\vec{D}}=\grad{\epsilon}\times\vec{E} + \epsilon(z)\curl{\vec{E}}
\end{equation}
Since $\vec{E}$ is an electrostatic field, $\curl{\vec{E}}=0$. Further, the electric field is parallel to the $z$ axis and so are the temperature and permittivity gradients. Therefore, the 
first term on the right hand side (RHS) of equation (\ref{e2}) is also zero, making $\vec{D}$ a conservative field. $\vec{D}$ can now be expressed as a gradient of a scalar field 
$\psi$. Note that $\psi(\vec{x})$ is \emph{not} the electric potential. If $\vec{D}=\grad{\psi}$, equation (\ref{e1}) gives $\Laplacian\psi=0$. The problem of finding $\vec{D}$ thus 
simplifies to solving Laplace's equation for $\psi$.

Since the plates are infinitely large, $\psi$ is a function of $z$ alone. The solution of $\Laplacian\psi=0$ is then $\psi=c_1z+c_2$, where $c_1$ and $c_2$ are constants of integration. 
The electric displacement is now $\vec{D}=c_1\uvec{z}$, where $\uvec{z}$ is a unit vector along the z axis. Since $\epsilon = \epsilon_0\kappa(z)$, where $\epsilon_0$ is the permittivity 
of free space and $\kappa$ is the relative permittivity, the electric field in the fluid is
\begin{equation}\label{e3}
\vec{E} = \frac{c_1}{\epsilon_0\kappa(z)}\uvec{z} = -\grad{\varphi},
\end{equation}
where $\varphi$ is the electric potential. The boundary conditions of the problem are given in terms of $\varphi$. They are 
\begin{eqnarray}
\varphi(z=L) &=& 0		\label{e4}	\\
\varphi(z=-L) &=&V		\label{e5}
\end{eqnarray}

Let $G$ be the temperature gradient, that is, $T(z)=T_0+Gz$ and $a$ be the coefficient of thermal variation of relative permittivity, that is, $\kappa(T)=\kappa_0+a(T-T_0)$
\cite{crc}. Using the form of $T(z)$, I get 
\begin{equation}\label{e6}
\kappa=\kappa_0+aGz
\end{equation}
Equation (\ref{e3}) then becomes
\begin{equation}\label{e7}
\frac{d\varphi}{dz} = \frac{-c_1}{\epsilon_0(\kappa_0+aGz)},
\end{equation}
the solution of which is 
\begin{equation}\label{e8}
\varphi(z) = \frac{-c_1}{aG\epsilon_0}\log\left|\frac{\kappa_0 + aGz}{c_3}\right|,
\end{equation}
where $c_3$ is another constant of integration. The two constants $c_1$ and $c_3$ are found from the experimental boundary conditions (\ref{e4}) and (\ref{e5}). They are 
\begin{eqnarray}
c_1 &=& -aG\epsilon_0 V\left\{\log\left|\frac{\kappa_0 - aGL}{\kappa_0 + aGL}\right|\right\}^{-1}	\label{e9} \\
c_3 &=& \kappa_0 + aGL	\label{e10}
\end{eqnarray}
The electric field itself is given by $\vec{E}=-\grad{\varphi}$, that is,
\begin{equation}\label{e11}
\vec{E}=\frac{E_0}{\kappa_0+aGz}\uvec{z} = \frac{E_0}{\kappa(z)}\uvec{z},
\end{equation}
where
\begin{equation}\label{e12}
E_0=\frac{c_1}{\epsilon_0} = -aGV\left\{\log\left|\frac{\kappa_0 - aGL}{\kappa_0 + aGL}\right|\right\}^{-1}
\end{equation}
$c_1$ being defined by equation (\ref{e9}). It is important to note that the electric field depends on $\kappa$, which in turn depends on the mass density through the Clausius-Mossotti
relation.

\section{Force on a volume element}\label{sec:vfd}
The force density inside a fluid dielectric is given, in SI units, by
\begin{equation}\label{e13}
\vec{f} = \rho_f\vec{E} - \frac{\epsilon_0 E^2}{2}\grad{\kappa} + \frac{\epsilon_0}{2}\grad{\left(\rho\frac{\partial}{\partial\rho}\left(\kappa E^2\right)\right)}
\end{equation}
It differs from the one given in \cite{pp} in that the electric field is assumed to depend on the mass density $\rho$. Refer to equation (71) of \cite{asj} for more details. From 
\eqref{e11}, $E^2 = E_0^2/\kappa^2(z)$ or $\kappa E^2 = E_0^2/\kappa$. Therefore,
\[
\frac{\partial}{\partial\rho}\left(\kappa E^2\right) = -\frac{E_0^2}{\kappa^2}\pdt{\kappa}{\rho}
\]
or
\[
\rho\frac{\partial}{\partial\rho}\left(\kappa E^2\right) = -\frac{E_0^2}{\kappa^2}\rho\pdt{\kappa}{\rho}
\]
The relation between mass density $\rho$ and electric permittivity $\kappa$ is provided by Clausius-Mossotti relation \cite{jdj}
\begin{equation}\label{e14}
\rho\frac{d\kappa}{d\rho}=\frac{\kappa^2+\kappa-2}{3}
\end{equation}
Therefore,
\[
\rho\frac{\partial}{\partial\rho}\left(\kappa E^2\right) = -\frac{E_0^2}{\kappa^2}\frac{\kappa^2+\kappa-2}{3} = -\frac{E_0^2}{3}\left(1 + \frac{1}{\kappa} - \frac{2}{\kappa^2}\right)
\]
and hence,
\[
\grad{\left(\rho\frac{d}{d\rho}(\kappa E^2)\right)} = -\frac{E_0^2}{3}\frac{4 - \kappa}{\kappa^3}\td{\kappa}{z}\uvec{z}
\]
From equation \eqref{e6},
\[
\grad{\left(\rho\frac{d}{d\rho}(\kappa E^2)\right)} = -aG\frac{E_0^2}{3}\frac{4 - \kappa}{\kappa^3}\uvec{z}
\]
or
\begin{equation}\label{e15}
\frac{\epsilon_0}{2}\grad{\left(\rho\frac{d}{d\rho}(\kappa E^2)\right)} = -aG\frac{\epsilon_0 E_0^2}{6}\frac{4 - \kappa}{\kappa^3}\uvec{z}
\end{equation}
Similarly,
\begin{equation}\label{e16}
\frac{\epsilon_0 E^2}{2}\grad{\kappa} = \frac{\epsilon_0}{2}\frac{E_0^2}{\kappa^2(z)}aG\uvec{z}
\end{equation}
Using equations \eqref{e15} and \eqref{e16} in \eqref{e13},
\begin{equation}\label{e17}
\vec{f} = -\epsilon_0aGE_0^2\left(\frac{2 + \kappa(z)}{3\kappa^3(z)}\right)\uvec{z}
\end{equation}
where we have used the fact that an ideal dielectric fluid is devoid of free charges, that is $\rho_f = 0$. 

The density variation due to temperature gradient is given by $\rho(z)=\rho_0e^{-\alpha Gz}$, where $\rho_0$ is the density of the fluid at the center, $z=0$ and $\alpha$ is the 
coefficient of volumetric thermal expansion. It is defined as
\begin{equation}\label{e18}
\alpha = \frac{1}{V}\left(\frac{\partial V}{\partial T}\right)_p = \frac{-1}{\rho}\left(\frac{\partial\rho}{\partial T}\right)_p,
\end{equation}
where $V$ is the volume, $p$ the pressure, $\rho$ the density and $T$ the temperature of the fluid. A subscript $p$ indicates pressure being held constant while evaluating partial 
derivatives. 

The total force on a volume element of density $\rho(z)$ is $\vec{F}(z) = \vec{f} - \rho(z)g\uvec{z}$ where $g$ in the second term is the acceleration due to gravity. Using 
equations (\ref{e12}), (\ref{e14}) and the dependence of density on $z$, I get the force on a volume element at $z$ 
\begin{equation}\label{e19}
\vec{F}(z) = -\left(\epsilon_0aGE_0^2\left(\frac{2 + \kappa(z)}{3\kappa^3(z)}\right) + \rho_0\exp(-\alpha Gz)g\right)\uvec{z}
\end{equation}
For most liquids $a < 0$ [p.6-166 to 6-187]\cite{crc}. Therefore, we write it as $a = -|a|$. Therefore, the above equation can be written as $\vec{F}(z) = F(z)\uvec{z}$, where
\begin{equation}\label{e20}
F(z) = \epsilon_0|a|GE_0^2\left(\frac{2 + \kappa(z)}{3\kappa^3(z)}\right) - \rho_0\exp(-\alpha Gz)g
\end{equation}
If, however $a > 0$, then $F(z)$ will always be negative and the net force on a volume element will always be downward.

\section{A criterion for onset of convection}\label{sec:crit}
Equation \eqref{e20} suggests that if there is no electric field, $F(z) < 0$, or $\vec{F}$ points downwards, for all $z \in [-L, L]$. Turning on the electric field, with an appropriately 
chosen voltage, one can make $F(z)$ positive in the interval $[-L, L]$. That is, it is possible to apply voltage across the capacitor so that the net force density on a volume element of 
the fluid points upwards at least in some portion of the fluid. Since $F$ is continuous in $[-L, L]$, it will flip its sign only if it vanishes at some point in the interval. To check if 
it vanishes once or multiple times, I first examine its first derivative. Let me write $F$ as
\begin{equation}\label{e21}
F(z) = A\left(\frac{2 + \kappa(z)}{\kappa^3(z)}\right) - Be^{-\alpha Gz}
\end{equation}
where
\begin{eqnarray}
A &=& \frac{\epsilon_0 |a| G E_0^2}{3} \label{e22} \\
B &=& \rho_0 g \label{e23}
\end{eqnarray}
Therefore,
\begin{equation}\label{e24}
\td{F}{z} = -2A\left(\frac{3 + \kappa(z)}{\kappa^4(z)}\right)aG + \alpha GB e^{-\alpha Gz}
\end{equation}
It will be zero, for if $E_0$ is estimated using the equation
\[
2A\left(\frac{3 + \kappa(z)}{\kappa^4(z)}\right)aG = \alpha GB e^{-\alpha Gz} \Rightarrow \frac{2}{3}\epsilon_0aE_0^2\left(\frac{3 + \kappa(z)}{\kappa^4(z)}\right) = \alpha Be^{-\alpha Gz}
\]
If we use ultra-pure water as the dielectric fluid, $\alpha = O(10^{-4})$, $\rho_0 = O(10^3)$, $\kappa = O(10^2)$, and $a = O(10^{-1})$. For the dimension of the capacitor, $z = 
O(10^{-3})$. Reasonable temperature gradients are a few tens of degrees per millimeter \cite{gp66} \cite{phr69}, which is $O(10^4)$. Therefore the right hand side of the above equation is 
$O(1)$. The left hand side can be estimated as $O(10^{-20})E_0^2$. Thus, the derivative of $F$ will be zero in the capacitor if $E_0$ is $O(10^{10})$ or voltage across the plates is 
$O(10^{12})$. Such a high electric field will cause dielectric breakdown [p.15-44]\cite{crc} and will, therefore, not be employed in an experiment. I can, thus assume that $dF/dz$ is never 
zero in the interval $[-L, L]$ or that, $F$ is monotonic throughout the capacitor cell.

If $F$ is monotonic in $[-L, L]$ and has to vanish at some point in the interval, $F(L)$ and $F(-L)$ will have opposite signs. For a typical field strength of $O(10)$ $kV/cm$\cite{gp66}, 
$dF/dz > 0$ and hence $F$ is monotone increasing, that is $F(L) > F(-L)$. Now, $F(L) > 0$ gives,
\begin{equation}\label{e25}
A\left(\frac{2 + \kappa_t}{\kappa_t^3}\right) > Be^{-\alpha GL}
\end{equation}
and $F(-L) < 0$ gives,
\begin{equation}\label{e26}
A\left(\frac{2 + \kappa_b}{\kappa_b^3}\right) < Be^{\alpha GL},
\end{equation}
where $\kappa_t = \kappa(L)$ is the permittivity at the top and $\kappa_b = \kappa(-L)$ is the permittivity at the bottom of the capacitor cell. From equations \eqref{e25} and \eqref{e26},
\[
B\left(\frac{\kappa_t^3}{2 + \kappa_t}\right)e^{-\alpha GL} < A < B\left(\frac{\kappa_b^3}{2 + \kappa_b}\right) e^{\alpha GL}
\]
or, using \eqref{e22}
\begin{equation}\label{e27}
\left(\frac{3B\kappa_t^3}{\epsilon_0|a|G(2 + \kappa_t)}\right)e^{-\alpha GL} < E_0^2 < \left(\frac{3B\kappa_b^3}{\epsilon_0|a|G(2 + \kappa_b)}\right) e^{\alpha GL}
\end{equation}
Thus, $E_0$ is between the limits $E_{min}$ and $E_{max}$, where
\begin{eqnarray}
E_{min} &=& \sqrt{\left(\frac{3B\kappa_t^3}{\epsilon_0|a|G(2 + \kappa_t)}\right)}\exp\left(-\frac{\alpha GL}{2}\right) \label{e28} \\
E_{max} &=& \sqrt{\left(\frac{3B\kappa_b^3}{\epsilon_0|a|G(2 + \kappa_b)}\right)}\exp\left( \frac{\alpha GL}{2}\right) \label{e29}
\end{eqnarray}
Using equation \eqref{e12}, I can conclude that if the voltage across the capacitor cell lies between $V_{min}$ and $V_{max}$, where
\begin{eqnarray}
V_{min} &=& \frac{1}{|a|G}\sqrt{\left(\frac{3B\kappa_t^3}{\epsilon_0|a|G(2 + \kappa_t)}\right)}\log\left|\frac{\kappa_b}{\kappa_t}\right|\exp\left(-\frac{\alpha GL}{2}\right)\label{e30} \\
V_{max} &=& \frac{1}{|a|G}\sqrt{\left(\frac{3B\kappa_b^3}{\epsilon_0|a|G(2 + \kappa_b)}\right)}\log\left|\frac{\kappa_b}{\kappa_t}\right|\exp\left( \frac{\alpha GL}{2}\right)\label{e31},
\end{eqnarray}
then $F(L)$ and $F(-L)$ will have opposite signs and there will be a region in the fluid where $F$ is zero. 

If the voltage across the capacitor cell is chosen between $V_{min}$ and $V_{max}$ then there is a region, $\mathcal{R}_1$, in the cell where $\vec{F}$ points upwards and the rest, 
$\mathcal{R}_2$,  where $\vec{F}$ points downwards. The two regions are separated by points where $\vec{F}$ vanishes. Consider a fluid parcel in $\mathcal{R}_2$. If a small perturbation
pushed it past the line where $\vec{F}$ vanishes, it ends up being in a region where $\vec{F}$ points upwards. Such a parcel, does not return to its original position. Similarly, a fluid
parcel, initially in $\mathcal{R}_1$, if perturbed to be in $\mathcal{R}_2$ fails to return to its original position. Thus, this choice of the voltage can trigger electrohydrodynamic 
convection.

\section{Estimating the voltage that can trigger convection}\label{sec:veri}
\begin{figure}[!h]
\centering
\centerline{\includegraphics[scale=0.5]{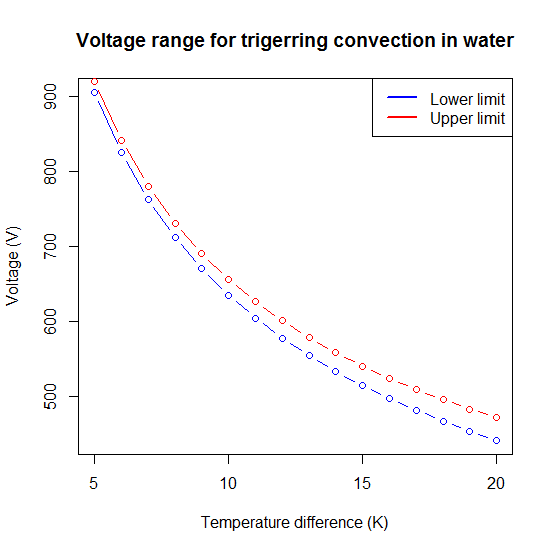}}
\caption{Water}
\label{veri:f1}
\end{figure}

I assume that a temperature difference of $15\,^{\circ}\mathrm{C}$ is applied across a gap of $1$ mm of ultra-pure water in a capacitor cell. Using the physical parameters in table 
\ref{t1}, in equations \eqref{e30} and \eqref{e31}, I get $V_{min} = 514.22$ $V$ and $V_{max} = 540.26$ $V$. This range of voltages is easily accessible in a laboratory and one
can devise a table-top experiment to verify the theory. Figure \ref{veri:f1} shows the range of voltages triggering convection in water for temperature differences between $5$ and $20$ 
$K$. It is important not to let conduction erase the temperature difference during the experiment. 
\begin{table}
\begin{center}
\begin{tabular}{lcr}
\bf{Parameter} 	& \bf{Description}					& \bf{Value}		\\[3 pt]
$\epsilon_0$	& Permittivity of free space 			& 8.85E-12	\\
L 		& Half-width of capacitor 			& 5E-4		\\
$\Delta T$	& Temperature difference between the plates	& 15		\\
a 		& Coeff. of temperature dependence of $\kappa$ 	& -0.79069		\\
$\kappa_0$ 	& Relative permittivity at $z=0$ 		& 249.21		\\
$\alpha$ 	& Coefficient of thermal expansion 		& 0.000214	\\
$\rho_0$ 	& Mass density at $z=0$ 			& 1e3		\\
$g$ 		& Acceleration due to gravity 			& 9.8		\\
\end{tabular}
\caption{Experimental parameters for water in SI units}
\label{t1}
\end{center}
\end{table}

\begin{figure}[!h]
\centering
\centerline{\includegraphics[scale=0.5]{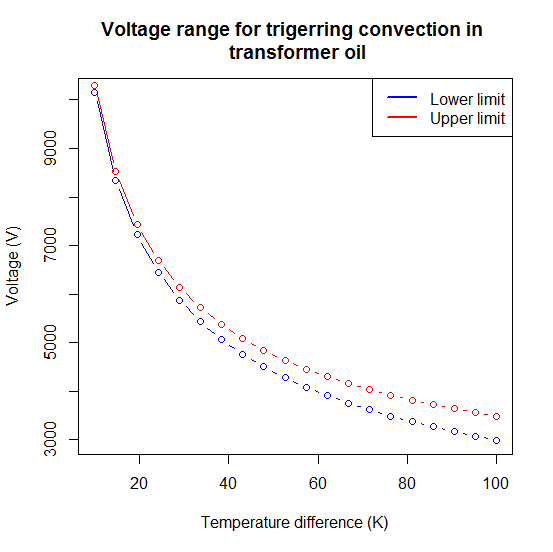}}
\caption{Transformer oil}
\label{veri:f2}
\end{figure}

If I carry out similar calculations for transformer oil, using parameters in table \ref{t2}, I get $V_{min} = 4403.06$ $V$ and $V_{max} = 4751.67.19$ $V$. These parameters were chosen from 
\cite{phr69}, except that:
\begin{itemize}
\item Coefficient of thermal expansion and density of transformer oil were obtained \cite{mat}
\item Relative permittivity was chosen so that the minimum value in the cell was not less than $1$.
\end{itemize}
The values of minimum and maximum voltage are an order of magnitude more than the ones reported in \cite{gp66} and \cite{phr69}. This could be either because the precise values of physical 
parameters are not known or the effect is because of free charges, as suspected in \cite{gp66}. The existence of free charges is a departure from dielectric behavior. It needs to be 
verified by an experiment if choosing transformer oil free of impurities shows convection at voltages lower than that predicted in this paper.
\begin{table}
\begin{center}
\begin{tabular}{lcr}
\bf{Parameter} 	& \bf{Description}					& \bf{Value}		\\[3 pt]
$\epsilon_0$	& Permittivity of free space 			& 8.85E-12	\\
L 		& Half-width of capacitor 			& 5E-4		\\
$\Delta T$	& Temperature difference between the plates	& 50		\\
a 		& Coeff. of temperature dependence of $\kappa$ 	& -1e-3		\\
$\kappa_0$ 	& Relative permittivity at $z=0$ 		& 1.2		\\
$\alpha$ 	& Coefficient of thermal expansion 		& 0.00086	\\
$\rho_0$ 	& Mass density at $z=0$ 			& 850		\\
$g$ 		& Acceleration due to gravity 			& 9.8		\\
\end{tabular}
\caption{Experimental parameters for transformer oil in SI units}
\label{t2}
\end{center}
\end{table}
A range of voltages triggering convection in transformer oil for temperature differences between $10$ and $100$ $K$ is shown in figure \ref{veri:f2}.
\bibliographystyle{amsplain}

\appendix
\section{Code to generate figures}
The R code to generate figure \ref{veri:f1} is given in listing \ref{list1}.
\lstset{basicstyle=\ttfamily\footnotesize, language=R, caption=Estimation of voltage range for water, label=list1}
\begin{lstlisting}[frame=single]
estimateV <- function(DeltaT) {
	# Permittivity of free space
	epsilon_0 <- 8.85e-12	
	
	# Half width of capacitor
	L <- 5e-4			
	
	# Coeff. of temperature dependence of permittivity
	a <- -0.79069		
	
	# Coeff. of thermal expansion
	alpha <- 0.000214	
		
	rho_0 <- 1000		# Mass density
	g <- 9.8		# Acceleration due to gravity
	k_0 <- 249.21		# Permittivity at z = 0

	G <- DeltaT/(2 * L)

	k_b <- k_0 - a*G*L
	k_t <- k_0 + a*G*L

	B <- 3*rho_0*g
# F(L) and F(-L) will have opposite sides if E_0^2 is between
	L1 <- (B*k_t^3*exp(-alpha*G*L))/(epsilon_0*abs(a)*G*(2+k_t))
	L2 <- (B*k_b^3*exp( alpha*G*L))/(epsilon_0*abs(a)*G*(2+k_b))
#or, E0 is between
	M1 <- sqrt(L1)
	M2 <- sqrt(L2)
#or, voltage is between
	V1 <- M1*log((k_0 - a*G*L)/(k_0 + a*G*L))/(abs(a)*G)
	V2 <- M2*log((k_0 - a*G*L)/(k_0 + a*G*L))/(abs(a)*G)

	return( list(V1, V2) )
}

tRange <- seq(from = 5, to = 20)
limits <- estimateV(tRange)

xl <- 'Temperature difference (K)'
yl <- 'Voltage (V)'
ml <- 'Voltage range for trigerring convection in water'
plot(tRange, limits[[1]], type = 'b', col = 'blue', 
	xlab = xl, ylab = yl, main = ml)
lines(tRange, limits[[2]], type = 'b', col = 'red')
legend.list <- c('Lower limit', 'Upper limit')
legend(x = 'topright', legend.list, lty = c(1, 1), 
	lwd=c(2.5,2.5), col=c('blue','red'))
\end{lstlisting}

Same code can be used to generate figure \ref{veri:f2} after changing the labels for the plot and choosing parameters as -
\lstset{basicstyle=\ttfamily\footnotesize, language=R, caption=Parameters for transformer oil, label=list2}
\begin{lstlisting}[frame=single]
	# Coeff. of temperature dependence of permittivity
	a <- -1e-3	
	
	# Coeff. of thermal expansion
	alpha <- 0.00086	
		
	rho_0 <- 840		# Mass density
	
	k_0 <- 1.2		# Permittivity at z = 0
\end{lstlisting}

\section{Version history}
\begin{enumerate}
\item First draft.
\item Corrected a spelling, added a line telling how this treatment varies from the previous ones and used correct order of magnitude of relative permittivity of water.
\end{enumerate}
\end{document}